\documentclass[lettersize,journal]{IEEEtran}

\usepackage[english]{babel}
\usepackage{blindtext}
\usepackage{subcaption}
\usepackage{url}
\usepackage[shortlabels]{enumitem}
\usepackage{mathtools} 
\usepackage{balance}
\usepackage[ruled,linesnumbered]{algorithm2e} %
\usepackage{natbib}
\usepackage[most]{tcolorbox}  
\usepackage[table]{xcolor}
\usepackage{booktabs}
\newcommand\systemname{\textsc{BShare}\xspace}

\usepackage{tikz}
\usepackage{xcolor}
 
\usetikzlibrary{arrows.meta}
 
\definecolor{outerbg}  {RGB}{235,235,235}
\definecolor{panelbg}  {RGB}{195,195,195}
\definecolor{panelbdr} {RGB}{130,130,130}
\definecolor{boxfill}  {RGB}{215,215,215}
\definecolor{c1}       {RGB}{ 49, 99,158}
\definecolor{c2}       {RGB}{204,130, 20}
\definecolor{c3}       {RGB}{160, 44, 44}
\definecolor{c4}       {RGB}{ 55,120, 80}

\begin{document}

\title{\systemname: Packet Queueing Delay-Driven Buffer Sharing for Datacenter Switches}


\author{\IEEEauthorblockN{Krishna Agarwal\IEEEauthorrefmark{1},
Muhamad Rizka Maulana\IEEEauthorrefmark{1},
Vamsi Addanki\IEEEauthorrefmark{2},
Habib Mostafaei\IEEEauthorrefmark{1}}

\IEEEauthorblockA{\IEEEauthorrefmark{1}Eindhoven University of Technology, The Netherlands}
\IEEEauthorblockA{\IEEEauthorrefmark{2}Purdue University, USA}
}

\maketitle
\begin{abstract}
Modern datacenter switches share packet buffers across ports to boost overall throughput and reduce packet loss. However, as buffer availability per-port-per-bandwidth unit continues to decrease, existing buffer-sharing strategies face increasing performance challenges. Recent efforts have attempted to integrate Buffer Management (BM) with Active Queue Management (AQM) to harness the advantages of both BM and AQM approaches to improve performance. While these hybrid solutions show promise, their complexity of dynamically calculating multiple factors for integration hinders generalization and efficiency.
This paper presents \systemname, a simple buffer sharing mechanism that uses packet queueing delay. \systemname requires only a single operator-configurable parameter. Our simulation results show that \systemname improves the flow completion time (FCT) performance of advanced transport protocols, such as PowerTCP, by up to 45.07\% compared to ABM, particularly under burst-heavy datacenter workloads.

\end{abstract}
\sloppy

\section{Introduction}

Datacenter switches rely on shared on-chip buffers to absorb traffic bursts and maintain high throughput under load. However, as switch capacities have increased, buffer sizes have not kept pace, primarily due to the rising cost and design complexity of high-speed memory. As a result, the amount of buffer available per-port per unit of bandwidth has steadily decreased. This trend is particularly problematic for datacenters, in which traffic is highly bursty, even at microsecond granularity~\cite{microscopic-IMC22}.

To cope with the limited buffer size of network switches, recent work has explored more intelligent buffer-sharing mechanisms that can dynamically allocate memory across ports and priorities to reduce packet loss during congestion~\cite{ShallowBuffer-ToN21,abm-sigcomm22,credence-nsdi24,l2bm-icdcs23}. Yet, this dynamic sharing introduces its challenges. When one queue grows aggressively, it can starve or throttle others, causing unfairness, throughput degradation, and tail latency inflation. Critically, such interference can occur even across logically independent queues, whether they originate from different applications or are mapped to separate output ports~\cite{credence-nsdi24}.

Datacenter switches rely on two primary mechanisms to manage packet admission and buffer allocation: Buffer Management (BM) and Active Queue Management (AQM). BM schemes such as Complete Sharing (CS)~\cite{completeSharing-ACMTA08}, Dynamic Threshold (DT)~\cite{DT-ToN98}, and Enhanced Dynamic Threshold (EDT)~\cite{EDT-TPDS17} aim to improve fairness in buffer utilization across output queues and reduce packet drop rates during congestion. These policies regulate buffer access to ensure that no single queue monopolizes shared memory, helping balance flow treatment under contention.
In contrast, AQM algorithms--including RED~\cite{RED-ToN93}, CoDel~\cite{CoDel}, and PIE~\cite{PIE-hpsr13}, operate by monitoring queue occupancy and proactively dropping packets before buffers overflow. This early drop behavior prevents persistent congestion and keeps queuing delays bounded, improving tail latency and responsiveness.
However, these two control schemes typically operate in isolation. While BM focuses on spatial fairness across queues, AQM targets temporal responsiveness within a single queue. Lack of coordination between them can lead to undesirable side effects such as queue starvation, misallocated buffer resources, and missed opportunities for holistic congestion control.

Recent proposals such as ABM~\cite{abm-sigcomm22}, L2BM~\cite{l2bm-icdcs23}, Reverie~\cite{Reverie-NSDI24}, and Credence~\cite{credence-nsdi24} have pushed the frontier of switch buffer sharing by combining BM with AQM, machine learning, or fine-grained telemetry. While effective in simulation or theoretical settings, these approaches rely on system components that are infeasible in today's programmable switches. For example, ABM and L2BM depend on the estimation of the per-queue drain rate and floating point arithmetic to dynamically adjust thresholds, but these drain rates must be periodically calculated using continuous metrics, often every 10--30 ms, which are highly sensitive to estimation noise, especially under shared link contention or bursty flows~\cite{PIE-hpsr13, prioMeter-euroP4}. Similarly, Reverie introduces ML-based prediction logic or oracle-assisted control paths, which, while conceptually powerful, assume access to operations (e.g., push-out, priority remapping, predictive feedback) not currently supported in line-rate switch hardware.

Moreover, many of these systems rely on DT to manage buffer sharing, which is known to induce buffer under-utilization in real deployments due to its conservative drop behavior~\cite{DT-ToN98}. These limitations raise a fundamental question.

\begin{tcolorbox}[colback=gray!10, colframe=gray!60!black, boxrule=0.5pt, arc=2pt, left=4pt, right=4pt, top=2pt, bottom=2pt]
Can we design a buffer sharing mechanism that is \emph{effective} under dynamic traffic conditions?
\end{tcolorbox}

\systemname answers this question by shifting complexity from the ingress path and instead uses \emph{queuing delay at the egress} as a congestion signal. This design has several advantages: (i) queuing delay is directly observable in hardware and reflects real-time congestion, even under high ingress arrival rates; and (ii) delay-based signals eliminate the need for per-queue drain rate estimation or traffic prediction. 
In doing so, \systemname remains simple and hardware-friendly, requiring only a single operator-configurable parameter, while matching or exceeding the performance of more complex designs.

Inspired by CoDel's delay-based philosophy, \systemname~\footnote{An early version of this idea appeared as a short position paper at a student workshop~\cite{CBM-CoNEXT-SW24}. This paper presents a complete system design and comprehensive evaluation that were not part of the earlier version.} dynamically shares switch buffer space across queues in datacenters, operating similarly to CS~\cite{completeSharing-ACMTA08} for buffer admission. Packets are dropped either when the buffer is full or when their queuing delay exceeds a target threshold that adapts to current buffer occupancy and congestion conditions. 
Our evaluation shows that \systemname performs comparably to ABM in most scenarios and outperforms ABM by up to 45.07\% in Flow Completion Time (FCT) when running modern transport protocols like PowerTCP~\cite{powertcp-nsdi22}.

Our contributions can be summarized as follows.
\begin{itemize}
    \item We propose \systemname, a lightweight buffer sharing mechanism that leverages queueing delay inspired by the CoDel AQM scheme to dynamically share buffer space across queues.
    
    \item We conduct extensive ns-3 simulations demonstrating that \systemname significantly improves flow completion time (FCT) slowdown compared to static buffer management (BM) schemes. Moreover, \systemname outperforms adaptive buffer management (ABM) approaches when combined with modern datacenter transport protocols.
    
    \item We perform a detailed steady-state analysis to characterize the behavior, stability, and buffer allocation dynamics of \systemname under varying traffic conditions.
\end{itemize}
 
\section{Background and Motivation}

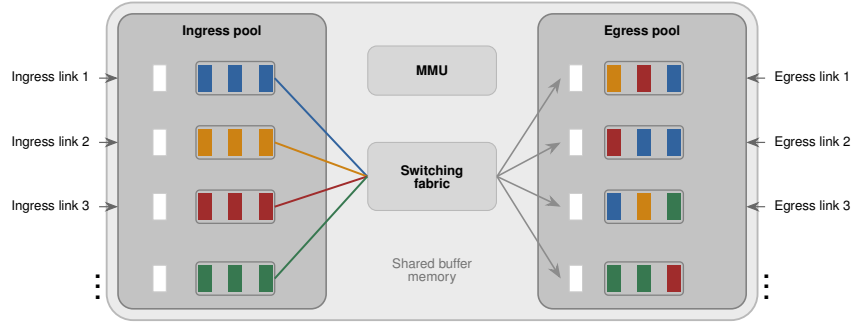
\begin{figure*}[tp]
    \centering
\begin{tikzpicture}[x=1mm, y=1mm,
  font=\sffamily,
  >=Stealth,
  cell/.style  = {minimum width=5pt, minimum height=10pt,
                  inner sep=0pt, outer sep=0pt},
  wcell/.style = {cell, fill=white, draw=gray!40, line width=0.2pt},
  cbox/.style  = {rounded corners=0.5mm, fill=none,
                  draw=panelbdr, line width=0.5pt},
  link/.style  = {->, line width=0.45pt, color=black!60},
  rlink/.style = {<-, line width=0.45pt, color=black!60},
]
 
\draw[rounded corners=3.5mm, fill=outerbg, draw=gray!55, line width=0.7pt]
  (0,0) rectangle (86,42);
 
\draw[rounded corners=2mm, fill=panelbg, draw=panelbdr, line width=0.6pt]
  (1.5,1.5) rectangle (29,40.5);
\node[font=\sffamily\bfseries\tiny, text=black] at (15.25,38.2) {Ingress pool};
 
\draw[rounded corners=2mm, fill=panelbg, draw=panelbdr, line width=0.6pt]
  (57,1.5) rectangle (84.5,40.5);
\node[font=\sffamily\bfseries\tiny, text=black] at (70.75,38.2) {Egress pool};
 
\node[draw=gray!55, fill=boxfill, rounded corners=1.2mm,
      minimum width=17mm, minimum height=6.5mm,
      font=\sffamily\bfseries\tiny] (MMU) at (43,33) {MMU};
 
\node[draw=gray!55, fill=boxfill, rounded corners=1.2mm,
      minimum width=17mm, minimum height=9mm,
      font=\sffamily\bfseries\tiny, align=center] (SF) at (43,19)
      {Switching\\[-1pt]fabric};
 
\node[font=\sffamily\tiny, align=center, text=black!55]
  at (43,6.5) {Shared buffer\\[-1pt]memory};
 
\def\rA{32}  \def\rB{23.5}  \def\rC{15}  \def\rD{5.5}
 
\def\cbxl{11.8}  \def\cbxr{22.2}  
\def\cbhh{2.2}                     
 
\def\ecbxl{65.8}  \def\ecbxr{76.2}
 
\foreach \ry/\col in {\rA/c1, \rB/c2, \rC/c3, \rD/c4}{
  \node[wcell]          at (7,\ry)   {};
  \draw[cbox] (\cbxl,{\ry-\cbhh}) rectangle (\cbxr,{\ry+\cbhh});
  \node[cell,fill=\col] at (13,\ry)  {};
  \node[cell,fill=\col] at (17,\ry)  {};
  \node[cell,fill=\col] at (21,\ry)  {};
}
 
\node[wcell]        at (62,\rA){}; \draw[cbox] (\ecbxl,{\rA-\cbhh}) rectangle (\ecbxr,{\rA+\cbhh});
\node[cell,fill=c2] at (67,\rA){}; \node[cell,fill=c3] at (71,\rA){}; \node[cell,fill=c1] at (75,\rA){};
\node[wcell]        at (62,\rB){}; \draw[cbox] (\ecbxl,{\rB-\cbhh}) rectangle (\ecbxr,{\rB+\cbhh});
\node[cell,fill=c3] at (67,\rB){}; \node[cell,fill=c1] at (71,\rB){}; \node[cell,fill=c1] at (75,\rB){};
\node[wcell]        at (62,\rC){}; \draw[cbox] (\ecbxl,{\rC-\cbhh}) rectangle (\ecbxr,{\rC+\cbhh});
\node[cell,fill=c1] at (67,\rC){}; \node[cell,fill=c2] at (71,\rC){}; \node[cell, fill=c4]        at (75,\rC){};
\node[wcell]        at (62,\rD){}; \draw[cbox] (\ecbxl,{\rD-\cbhh}) rectangle (\ecbxr,{\rD+\cbhh});
\node[cell,fill=c4] at (67,\rD){}; \node[cell,fill=c4] at (71,\rD){}; \node[cell,fill=c3] at (75,\rD){};
 
\draw[c1, line width=0.75pt] (22.2,\rA) -- (SF.west);
\draw[c2, line width=0.75pt] (22.2,\rB) -- (SF.west);
\draw[c3, line width=0.75pt] (22.2,\rC) -- (SF.west);
\draw[c4, line width=0.75pt] (22.2,\rD) -- (SF.west);
 
\draw[->, black!45, line width=0.6pt] (SF.east) -- (60,\rA);
\draw[->, black!45, line width=0.6pt] (SF.east) -- (60,\rB);
\draw[->, black!45, line width=0.6pt] (SF.east) -- (60,\rC);
\draw[->, black!45, line width=0.6pt] (SF.east) -- (60,\rD);
 
\node[font=\sffamily\tiny, anchor=east] at (-1,\rA) {Ingress link 1};
\node[font=\sffamily\tiny, anchor=east] at (-1,\rB) {Ingress link 2};
\node[font=\sffamily\tiny, anchor=east] at (-1,\rC) {Ingress link 3};
\node[font=\sffamily\bfseries\small, anchor=east] at (0.5,\rD) {$\vdots$};
\draw[link] (-1,\rA) -- (1.5,\rA);
\draw[link] (-1,\rB) -- (1.5,\rB);
\draw[link] (-1,\rC) -- (1.5,\rC);
 
\node[font=\sffamily\tiny, anchor=west] at (87,\rA) {Egress link 1};
\node[font=\sffamily\tiny, anchor=west] at (87,\rB) {Egress link 2};
\node[font=\sffamily\tiny, anchor=west] at (87,\rC) {Egress link 3};
\node[font=\sffamily\bfseries\small, anchor=west] at (85.5,\rD) {$\vdots$};
\draw[rlink] (84.5,\rA) -- (87,\rA);
\draw[rlink] (84.5,\rB) -- (87,\rB);
\draw[rlink] (84.5,\rC) -- (87,\rC);

\end{tikzpicture}
    \caption{The architecture of a shared-memory switch in which different colours indicate different sources and priorities of data packets with one queue per output link.}
    \label{fig:SwitchArchitecture}
\end{figure*}

\subsection{Background}\label{2:Switching Complications}

A shared memory switch architecture is a widely adopted design in high-performance networking devices, as illustrated in Figure~\ref{fig:SwitchArchitecture}, including switches and routers, to efficiently manage network packet flow. 

A central feature of this architecture is a buffer memory accessible by multiple components within the switch. Incoming packets at input ports are directed to ingress queues for temporary storage, which regulates entry into the switch fabric. Egress queues at output ports similarly store packets before transmission onto the outgoing network. BM algorithms are employed to allocate buffer space efficiently among these queues. The shared buffer memory is a critical resource, providing rapid access and storage for packets from multiple ingress and egress points simultaneously. This architecture typically incorporates a memory management unit (MMU) that arbitrates access to the shared memory, ensures fairness through BM and AQM, and optimizes throughput. Leveraging shared memory resources enables the switch to accommodate fluctuating traffic loads and maintain high throughput, a crucial aspect of modern network performance.

Switches play a central role in forwarding packets across modern networks. However, packet forwarding within switches faces several challenges, including congestion, buffer contention, and inefficient queue management, all of which can significantly impact network performance. In this section, we discuss the key factors affecting switch performance, with a particular focus on congestion, starvation, and queueing delay, and analyze their impact on packet forwarding efficiency and application performance.

\begin{itemize}
\item \textbf{Congestion.} Congestion in switch queues refers to a situation where the amount of incoming traffic exceeds the capacity of the switch's output ports or the processing capabilities of the switch itself.

\noindent \textbf{Effect.} It results in delays in packet forwarding, increased queuing delay, packet loss, and overall degradation of network performance~\cite{ComputerNetworks}. 

\item \textbf{Queuing Delay.} When packets arrive at a network device, such as a switch or router, they are stored in queues before being processed and forwarded. Queuing delay is the time a packet spends waiting in these queues before being transmitted. As network traffic increases and queues become longer, the queuing delay also increases. 

\noindent \textbf{Effect.} This delay can impact real-time applications such as video streaming or VoIP, causing jitter and latency issues. The other term, which is often heard and related to queueing delay, is \textbf{Bufferbloat}, which occurs when a network device has large buffers. Having a large buffer means it can store a large number of packets, but it leads to increased latency and delays in delivering packets ~\cite{ComputerNetworks}.

\item \textbf{Starvation.} Starvation arises when specific flows or types of traffic consistently receive lower priority or insufficient network resources compared to others.

\noindent \textbf{Effect.} This disparity in resource allocation leads to poor performance for the starved traffic. For instance, if a network prioritizes certain applications over others, lower-priority applications may experience persistent delays or insufficient bandwidth, resulting in a state of starvation ~\cite{abm-sigcomm22}.

\end{itemize}

\subsection{Motivation}
In this section, we present the limitations of current switch buffer sharing algorithms and then motivate the need to design our solution.

\noindent\textbf{Limitations of existing designs.}
As datacenter switch buffers continue to shrink and traffic becomes increasingly bursty and unpredictable, buffer sharing mechanisms must evolve to remain practical and deployable. Recent proposals, such as ABM~\cite{abm-sigcomm22}, improve switch buffer sharing by combining spatial (BM) and temporal (AQM) control or by leveraging optimization-based approaches.  While these designs demonstrate strong performance in simulation, they rely on calculating drain rates for each queue. The calculation of drain rates involves determining continuous values, which are susceptible to estimation errors, especially when multiple queues share the same link or link capacity fluctuates~\cite{PIE-hpsr13, prioMeter-euroP4}.

We argue that these designs have two key limitations:

(i) \textbf{Unrealistic assumptions}: These schemes often depend on idealized switch behavior. For example, they require fine-grained per-queue metrics such as instantaneous drain rate, which are difficult or impossible to measure accurately in hardware. Some techniques assume the availability of unsupported features like push-out, floating-point operations, or solver-based coordination.

(ii) \textbf{Implementation complexity}: Even if theoretically implementable, these designs require tightly coupled control across multiple queues and memory regions. For example, ABM depends on continuously updated per-queue drain rates and coordinates global buffer state to make drop decisions, logic that is difficult to parallelize and incompatible with the match-action abstraction used in programmable switches.

Therefore, we design a practical buffer sharing mechanism that avoids reliance on carefully tuned parameters while achieving performance comparable to prior buffer sharing schemes.

\section{System Design}

We aim to create a single buffer-sharing algorithm that addresses the limitations of existing methods while incorporating the following desired features:
(i) \textit{Simple to configure}, \systemname uses a single operator-configurable parameter (\(\alpha_p\)) per priority class.
(ii) \textit{Driven by queuing delay}, rather than raw queue lengths, enabling more accurate congestion response and fair buffer allocation.

\systemname's core idea is that packet queuing delay, a coarse but powerful signal, can serve as a scalable proxy for drain time and congestion, without requiring heavyweight calculations or coordination. The threshold for each queue is dynamically adjusted using local delay observations and minimal shared-state arithmetic.

\subsection{The \systemname Algorithm}
\systemname\ draws inspiration from the simplicity and effectiveness of packet queueing delay of CoDel~\cite{CoDel} and the spatial efficiency of CS~\cite{completeSharing-ACMTA08}. In an output-queued shared-memory packet switching chip, \systemname assigns a threshold $\theta_p^i(t)$ for dequeuing a packet with priority $p$ and port $i$ for any particular instance of time $t$. Here, $p$ represents the service priority of the packet, distinguishing traffic classes such as short flows, incast bursts, or background flows, each of which may tolerate different levels of queuing delay.
This threshold enables \systemname\ to adaptively regulate buffer usage based on real-time congestion signals while maintaining hardware implementability.

Assume that $\mathcal{B}$ denotes the total available buffer space in the switch, and $Q(t)$ represents the amount of buffer used to enqueue current packets across all ports and queues. \systemname computes the delay threshold $\theta_p^i(t)$ using a configurable parameter $\alpha_p$, the port bandwidth $\mathcal{C}$, and two dynamic system metrics:
(i) the number of congested queues at priority level $p$, denoted by $c_p$, and
(ii) the available buffer space, given by $\mathcal{B} - Q(t)$.
The delay threshold is then defined as:

\begin{equation}\label{eq:Codel} \theta_p^i(t) = \frac{\alpha_p \cdot (\mathcal{B} - Q(t))}{c_p \cdot \mathcal{C}} \end{equation}

\noindent \colorbox{lightgray}{$\alpha_p$} is the only operator-configurable parameter. It scales the aggressiveness of delay adaptation: a larger $\alpha_p$ increases a queue’s chance of accepting packets under congestion, similar to the role of the delay target in DT~\cite{DT-ToN98}. This flexibility allows tailoring queue behavior based on service-level requirements or traffic priorities.

\noindent  \colorbox{lightgray}{$C$}  denotes the port bandwidth. A higher bandwidth implies faster draining of queues, allowing less time to hold packets before dequeuing. As a result, the delay threshold is inversely proportional to $C$, ports with higher bandwidths receive proportionally smaller delay thresholds to maintain timely queue management.

\noindent  \colorbox{lightgray}{$(\mathcal{B}-Q(t))$} captures the available buffer space at time $t$, representing how much of the buffer is currently unused. It reflects the system's willingness to admit more packets: the more free space, the more permissive the threshold. Similar to DT~\cite{DT-ToN98}, this design choice for available buffer space ensures that queues with more headroom are less aggressive in dropping packets. For example, when the buffer is half full, the threshold is twice as large as when it is nearly full, encouraging smoother utilization under varying load.

\noindent  \colorbox{lightgray}{$c_{p}$} denotes  number of congested queues at priority level $p$. A queue is considered congested when its length nears its threshold. In \systemname, we classify a queue as congested when its length reaches at least 90\% of the overall buffer space. The threshold is inversely related to $c_p$: as more queues become congested, the delay budget is divided among them, reducing the threshold and making each queue more conservative in accepting packets.

\subsection{Operational Overview}\label{sec:how-it-works}
We now present \systemname and explain how it works.
Upon arrival at the output port, \systemname performs an admission check before enqueueing the packet. Specifically, it verifies whether accepting the packet would exceed the switch's remaining buffer capacity $\mathcal{B} - Q(t)$. This check prevents buffer overflow and ensures stable queue behavior.

Following a similar philosophy to CS~\cite{completeSharing-ACMTA08}, \systemname compares the current buffer occupancy with the total available buffer of the switch. If sufficient space is available, the packet is admitted to the queue while retaining its original priority. Additionally, \systemname records the packet's arrival timestamp during dequeuing.
If the buffer is full, \systemname immediately drops the packet to avoid further congestion, maintaining backpressure and fairness across competing queues.

When packets accumulate in output (egress) queues, \systemname applies a delay-aware dequeuing policy to determine whether a packet should be transmitted or dropped. This decision is based on two metrics: (i) the packet's sojourn time, i.e., the time spent in the queue since arrival, and (ii) a dynamically computed target delay threshold, $\theta_p^i(t)$, as defined in Eq.~\ref{eq:Codel}.
If the sojourn time of a packet is below the target delay, the packet is forwarded for transmission. Otherwise, it is dropped, signaling excessive queuing and potential congestion. This delay-aware dropping mechanism ensures that stale packets do not occupy limited buffer space at the expense of newer, potentially more latency-sensitive traffic.

Datacenter traffic is often bursty at sub-RTT timescales, particularly during the initial phases of a connection when congestion control has not yet reacted. To protect short flows, which are highly sensitive to even a single packet drop, \systemname assigns higher priority (via a larger $\alpha$) to packets within their first RTT. This reduces their likelihood of being dropped under transient congestion events, such as incast bursts.

In contrast, lower priority traffic, such as long-lived flows, is assigned a smaller $\alpha$, resulting in a smaller target delay threshold. Packets from these flows that persist too long in the buffer are more likely to be dropped, enabling TCP to respond with backoff and alleviating pressure on shared buffers. This design balances throughput efficiency with latency sensitivity across diverse flow types.

\begin{algorithm}[t]
    \caption{\systemname algorithm}
    \label{alg:dequeue}
    \KwIn{None}
    \KwOut{packet - Packet to transmit.}
    \tcc{Perform round-robin scheduling over all queues}
    \ForEach{\texttt{queue} in \texttt{Queues}}{
        \texttt{packet} $\gets$ \texttt{queue.dequeue()}\;
        \If{packet $\neq$ null}{
            \texttt{priority} $\gets$ \texttt{packet.getPriority()}\;
            \texttt{targetDelay} $\gets \theta_p^i(t)$\;
            \tcc{sojournTime is the total time spent by packet in queue}
            \texttt{sojournTime} $\gets$ \texttt{CurrentTime()} $-$ \texttt{packet.getTimestamp()}\;
            \If{\texttt{sojournTime} $<$ \texttt{targetDelay} or \texttt{QueueLength(priority)} $<$ \texttt{MinBytes}}{
                \tcc{Admit packet for transmission}
                \Return packet\;
            }
            \Else{ 
                \tcc{Drop the packet and continue}
                DropAfterDequeue(packet)\;
            }
        }
    }
    \Return packet\;
\end{algorithm}

\noindent\textbf{Threshold update.}
In \systemname, the delay threshold $\theta_p^i(t)$, defined in Eq.~\ref{eq:Codel}, is updated upon every packet departure. This ensures threshold decisions remain responsive to real-time buffer dynamics while avoiding excessive update overhead.

The parameters $\alpha_p$ (priority weight), $\mathcal{B}$ (total buffer capacity), and $\mathcal{C}$ (port bandwidth) are static during runtime, offering stability and reducing tuning complexity.  However, the buffer size $Q(t)$ (used buffer space), is subject to continuous change after every packet enqueue and dequeue operation. 
The parameter $c_p$, denoting the number of congested queues at priority level $p$, is updated less frequently, once per RTT. This amortized update strategy provides a coarse but effective congestion signal without incurring per-packet processing overhead.

\noindent\textbf{Safeguard.}
\systemname enforces boundary conditions on buffer occupancy to maintain robust and predictable behavior under varying load. These safeguards prevent premature drops and uncontrolled buffer growth. If the queue length falls below a predefined minimum threshold (\texttt{MinBytes}), \systemname suppresses packet drops—regardless of queuing delay or priority. This ensures that buffers are not underutilized, avoiding unnecessary packet loss when sufficient capacity is available.
Conversely, if the buffer occupancy approaches its maximum capacity, \systemname immediately drops incoming packets to prevent overflow. This early rejection serves as a backpressure signal, helping to mitigate congestion and preserving stability across competing queues.
Together, these boundary checks ensure that \systemname remains efficient and resilient under both light and heavy traffic conditions.

\subsubsection*{Relation to CoDel}
While \systemname draws conceptual inspiration from CoDel's use of queuing delay as a congestion signal, the two serve fundamentally different roles. CoDel is an AQM algorithm designed to operate within a single queue, aiming to control latency by proactively dropping packets based on sojourn time and fixed control intervals. It uses a static target delay (typically 5ms) and adjusts its drop rate using an inverse square-root backoff mechanism over 100ms intervals.

In contrast, \systemname is a buffer sharing mechanism designed for switches with multiple output queues and shared on-chip memory. It does not attempt to manage congestion within a single queue but instead determines buffer admission and packet drop decisions across queues, using a dynamic, delay-based threshold. Unlike CoDel, \systemname requires no interval tracking or static parameters, and it is explicitly designed for efficient implementation in programmable switch data planes.

\noindent\systemname addresses several key limitations of CoDel's packet-dropping strategy:

\begin{enumerate}[(i)] \item \textbf{Dynamic threshold adaptation:} Unlike CoDel's fixed delay target, \systemname computes its delay threshold based on current buffer occupancy and queue contention, allowing it to adapt to varying traffic conditions in real time. \item \textbf{Simplified control logic:} CoDel requires interval tracking and square-root-based drop pacing. \systemname avoids this complexity, using a stateless, threshold-based check on every dequeued packet. 
\item \textbf{Faster congestion reaction:} CoDel may react slowly to short bursts due to its interval-based design~\cite{codelDrawback-lcn14}. \systemname enables immediate drop decisions based on per-packet delay, making it more responsive under bursty datacenter traffic. 
\end{enumerate}

CoDel and \systemname serve different purposes, AQM versus buffer management, so we will exclude CoDel in our evaluation. They cannot be interchanged in deployment, and comparing them would confuse congestion control within a queue with buffer allocation across queues.

\section{Steady State Analysis}\label{sec:stedayState}
In this section, we analyze the steady state of \systemname inspired by ABM~\cite{abm-sigcomm22}. 
Before checking for a steady state, let us consider the model and formalize its allocation.

\subsection{Formalizing \systemname allocation}
In our model, we consider a switch with a fixed number of ports and one queue per priority per port. The switch operates on a shared memory architecture with a total buffer space of $\mathcal{B}$. At any given time $t$, the occupied buffer space is denoted as $Q(t)$. Our analysis relies on a fluid model, assuming deterministic and continuous arrivals and departures of packet bits.

For \systemname, we use the parameter $\alpha_{p}$ to allocate buffer space for each priority $p$. Each priority is associated with a dedicated queue at each port. Port indices are denoted by $i$, and $p$ represents the priority. The set of priorities utilizing the buffer is denoted as $\wp$. The number of congested queues for a priority $p$ at time $t$ is represented by $c_{p}(t)$.

As described in the Equation~\ref{eq:Codel}, the threshold time of a packet during dequeue time at the port $i$ and belonging to priority $p$ is determined by the alpha parameter $\alpha_{p}$, the number of congested queues $c_{p}(t)$, port bandwidth $C$, and the remaining buffer space $\mathcal{B}- Q(t)$. This calculation can be formally expressed as:

\begin{equation}\label{eq:formal}
\Omega_{\text{deq},p}^{i}(t) = \frac{\alpha _{p} . \lambda_{p}(t) . (\mathcal{B}-Q(t)) } {C}
\end{equation}
where $\lambda_{p}(t) = \frac{1}{c_{p}(t)}$ is the inverse of the total number of congested queues of priority $p$ at time $t$.
We are introducing a new parameter delta $\Delta_{p}^{i}$ which is the product of $\alpha _{p}$ and $\lambda_{p}(t)$ having priority $p$ at port $i$ at instance time $t$.

\begin{equation}\label{eq:delta}
\Delta_{p}^{i}(t) = \alpha _{p} . \lambda_{p}(t)
\end{equation}

\subsection{Property of Delta}
The total instantaneous sum of $\Delta_{p}^{i}(t)$ across all queues within a priority $p \in \wp$ across all ports is constrained by an upper bound denoted as $\alpha_p$.

\begin{equation}\label{eq:sum of delta}
 \sum_{i}\Delta_{p}^{i}(t) \leq \alpha _{p}
\end{equation}

\textbf{Proof:} Observing that $\lambda_{p}(t)$ represents the number of congested queues for a priority $p$, it remains constant across all queues of the same priority; therefore, the equation is as follows:

\begin{equation}\label{eq:proof of delta}
 \sum_{i}\Delta_{p}^{i}(t) = \sum_{i}\alpha _{p}.\lambda_{p}(t) = \sum_{i}\alpha _{p}.\frac{1}{c_{p}(t)} = \alpha _{p}.\frac{c_{p}(t)}{c_{p}(t)} = \alpha _{p}
\end{equation}

\subsection{Analysis}
In the \systemname context, a steady state refers to a situation where the load conditions and buffer occupancy remain constant. Essentially, this means that the number of packets entering the system equals the number of packets leaving it, maintaining a balance. In a steady state, each packet has a threshold time to leave the system, which represents the ideal time after which the packet will be discarded if it remains in the buffer. During a steady state, these ideal times for packets remain consistent.

During the steady state of \systemname, our primary concern is to evaluate important metrics. These include the overall buffer allocation, denoted as $Q$ and represented by Equation~\ref{eq:buffer allocation}. Additionally, we are interested in determining the remaining buffer space, denoted as $\mathcal{B}-Q$, which is calculated using Equation~\ref{eq:remaining buffer space}. Moreover, we focus on calculating the thresholds per congested queue, represented as $\Omega_{\text{deq},p}^{i}$, as specified by Equation~\ref{eq:thresholds per congested queue}, using \systemname.

\begin{equation}\label{eq:buffer allocation}
Q =  \frac{\mathcal{B}\sum_i\sum_p \Delta_{p}^{i} } {1 + \sum_i\sum_p\Delta_{p}^{i}}
\end{equation}

\begin{equation}\label{eq:remaining buffer space}
\mathcal{B}-Q = \frac{B} {1 + \sum_i\sum_p\Delta_{p}^{i}}
\end{equation}

\begin{equation}\label{eq:thresholds per congested queue}
\Omega_{\text{deq},p}^{i} = \frac{B .\Delta_{p}^{i} } {(1 + \sum_i\sum_p\Delta_{p}^{i})}
\end{equation}

\noindent \textbf{Proof 1:} In the steady-state scenario, where we assume that the packet ideal time in each queue is equal to their thresholds, we can derive the overall buffer occupancy by summing the queue lengths of all congested queues of all ports and priorities. This can be represented as follows:

\begin{equation}\label{eq:proof buffer allocation}
\frac{Q}{C} =\sum_i\sum_p \frac{\Delta_{p}^{i} . (\mathcal{B}-Q)} {C}
\end{equation}

Solving this equation for $q$ leads to the final equality, which encapsulates the determination of overall buffer occupancy.\\

\noindent \textbf{Proof 2:} For remaining buffer allocation, we subtract Equation~\ref{eq:buffer allocation} from total buffer ($\mathcal{B}$).\\

\noindent \textbf{Proof 3:} To calculate thresholds per congested queue, we put the value of the remaining buffer from Equation~\ref{eq:remaining buffer space} in the below threshold calculation of \systemname and multiply by bandwidth (b) to convert to the threshold length of the queue.

\begin{equation}\label{eq:proof of thresholds per congested queue}
\Omega_{\text{deq},p}^{i} = \frac{\alpha _{p} (\mathcal{B}-Q).C} {c_{p} C}
\end{equation}

\subsection{Reduced dependency on scheduling algorithms}
The reduction of dependency on scheduling algorithms is highlighted in Equation~\ref{eq:Codel}. It is important to note that this equation does not rely on factors such as link rates, traffic loads, queue length, or drain rate. This reduces reliance on variables that are beyond the control or prediction of the local buffer. In a recent approach proposed by~\cite{abm-sigcomm22}, however, the drain rate is incorporated into the calculation. This introduces complexities as determining drain rates involves continuous value estimation over time, which can be prone to errors, especially in scenarios where multiple queues share the same link or when link capacity fluctuates~\cite{PIE-hpsr13, prioMeter-euroP4}.

\subsection{Minimal Buffer Waste}

In \systemname, we do not impose strict boundaries like \textit{CS} for buffers, which allows the utilization of the entire buffer capacity while maintaining fairness among queues, as observed in previous studies~\cite{CS1, CS2}, but this differs from \textit{DT}~\cite{DT-ToN98} and \textit{ABM}~\cite{abm-sigcomm22}, which allocate buffer space for transient conditions and bursty cases, resulting in potential underutilization when there are no bursty scenarios. However, \textit{CS} can lead to drawbacks, particularly when long flows occupy the buffer completely and cause packet drops for shorter flows, resulting in poor performance in terms of \textit{FCT} slowdown~\cite{FB}. Therefore, \textit{\systemname} strikes the right balance by prioritizing short flows and bursts dynamically without sacrificing buffer underutilization, considering time as a factor. \systemname makes sure of full buffer utilization, but in case the size of the buffer is not a multiple of packet size, there is a slight amount of buffer remaining, which is less than the packet size.

\subsection{\systemname: Assessing Property Alignment}
In this section, we analyse the extent to which \systemname adheres to the principles of ABM properties~\cite{abm-sigcomm22}.
\begin{itemize}
   \item \textbf{Isolation} 
Drawing from \textit{ABM}'s emphasis on isolation across system priorities, \systemname similarly prioritizes equitable resource distribution through tailored mechanisms. \systemname achieves this by (i) imposing constraints on the total buffer allocation for each priority and (ii) guaranteeing minimum buffer allocations to individual priorities. This ensures that no single priority can monopolize the buffer at the expense of others. \systemname dynamically adjusts thresholds per queue based on congestion levels within each priority, mirroring \textit{ABM}'s approach. As congestion increases within specific priority queues, \systemname reduces thresholds for each queue within that priority, thereby promoting fair resource allocation across priorities.

By applying the concept of isolation to \systemname, it remains effective, as evidenced by the equations of~\ref{eq:Isolation - Minimum guarantee} and~\ref{eq:Isolation - Preventing monopoly}. The equations below affirm \systemname's commitment to ensuring equitable resource distribution among system priorities.

\textbf{Minimum guarantee buffer} The minimum available buffer size for any priority level $p$ can be expressed as:

\begin{equation}\label{eq:Isolation - Minimum guarantee}
     \text{Minimum Buffer} \geq  \frac{\mathcal{B} .\alpha_{p}} {(1 + \sum_{p \in \wp }\alpha_{p})}
\end{equation}
    \noindent \textbf{Proof:} The threshold $\Omega_{\text{deq},p}^{i}$ for each queue with priority $p$ is determined by Equation~\ref{eq:thresholds per congested queue}. Summing across all ports, we compute the total allocated buffer as follows:
    
    \begin{equation}\label{eq:proof Isolation - Minimum guarantee}
    \sum_i\Omega_{\text{deq},p}^{i} = \frac{\mathcal{B} .\sum_i\Delta_{p}^{i} } {(1 + \sum_i\sum_{p \in \wp }\Delta_{p}^{i})} \geq  \frac{\mathcal{B} .\alpha_{p}} {(1 + \sum_{p \in \wp }\alpha_{p})}
    \end{equation}
   The last inequality holds since $\sum_{i}\sum_{p}\Delta_{p}^{i} \leq \sum_{p}\alpha _{p}$ from Equation~\ref{eq:sum of delta}.\\

\textbf{Buffer preventing monopoly} The maximum available buffer size for any priority level $p$ can be expressed as:

\begin{equation}\label{eq:Isolation - Preventing monopoly}
     \text{Maximum Buffer} \leq  \frac{\mathcal{B} .\alpha_{p}} {(1 + \alpha_{p})}
\end{equation}
\noindent \textbf{Proof:} The proof follows a similar approach to Theorem~\ref{eq:proof Isolation - Minimum guarantee}. To establish the upper bound, we leverage the property $\sum_{i}\sum_{p \in \wp }\Delta_{p}^{i} \geq \alpha _{p}$, which occurs when only priority $p$ is utilizing the buffer.

\item \textbf{Bounded drain time.}
\systemname effectively manages drain time by allocating buffer space in proportion to the drain rate of individual queues. This mechanism ensures that \systemname limits the queuing delay and the overall buffer drain time, regardless of the number of congested queues or the scheduling policy in use.

The thresholds established by \systemname provide upper bounds for the drain time $\tau$ of any queue with priority $p$, as given by:
   
\begin{equation}\label{eq:Bounded Drain Time}
     \tau \leq  \frac{\mathcal{B} .\alpha_{p}} {C (1 + \alpha_{p})}
\end{equation}

It should be noted that the upper bound of $\tau$ is only dependent on the constant parameters.

\noindent \textbf{Proof:} Using Equation~\ref{eq:thresholds per congested queue} from our steady-state analysis and considering that drain time is the occupied buffer divided by its bandwidth ($C$), we derive the drain time $\tau$ as follows for a queue at port $i$ and of priority $p$

\begin{equation}\label{eq:proof Bounded Drain Time}
     \tau =  \frac{\mathcal{B} .\alpha_{p}. \frac{1}{c_{p}}} {C(1 + \sum_i\sum_p\Delta_{p}^{i})} \leq \frac{\mathcal{B} .\alpha_{p}}{C(1 + \alpha_{p})}
\end{equation}
The last inequality holds since $\sum_i\sum_p\Delta_{p}^{i} \geq \alpha_{p}$ and $\frac{1}{c_{p}} \leq 1 $

\end{itemize}

\section{Performance Evaluation}\label{sec:eval}

\begin{figure*}[tp]
    \centering    
    \begin{subfigure}[b]{.24\linewidth}
        \centering
        \includegraphics[width=\linewidth]{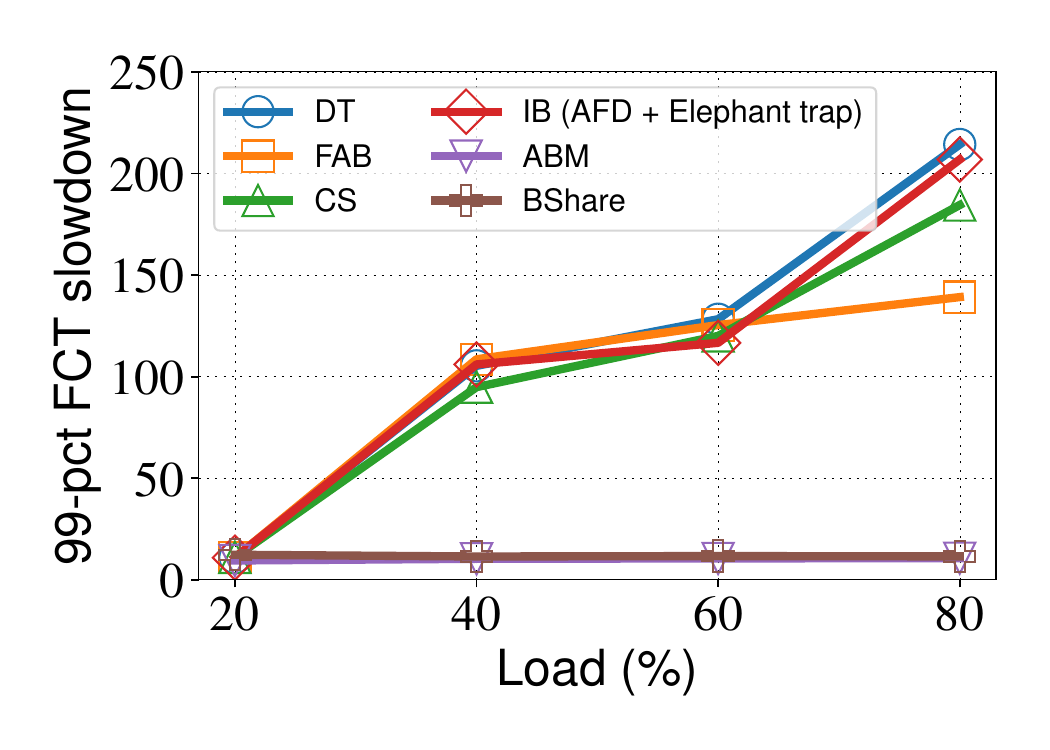}
        \caption{\texttt{Incast Flows}}
        \label{fig:load:Incast}
    \end{subfigure}    
    \hfill
    \begin{subfigure}[b]{.24\linewidth}
        \centering
        \includegraphics[width=\linewidth]{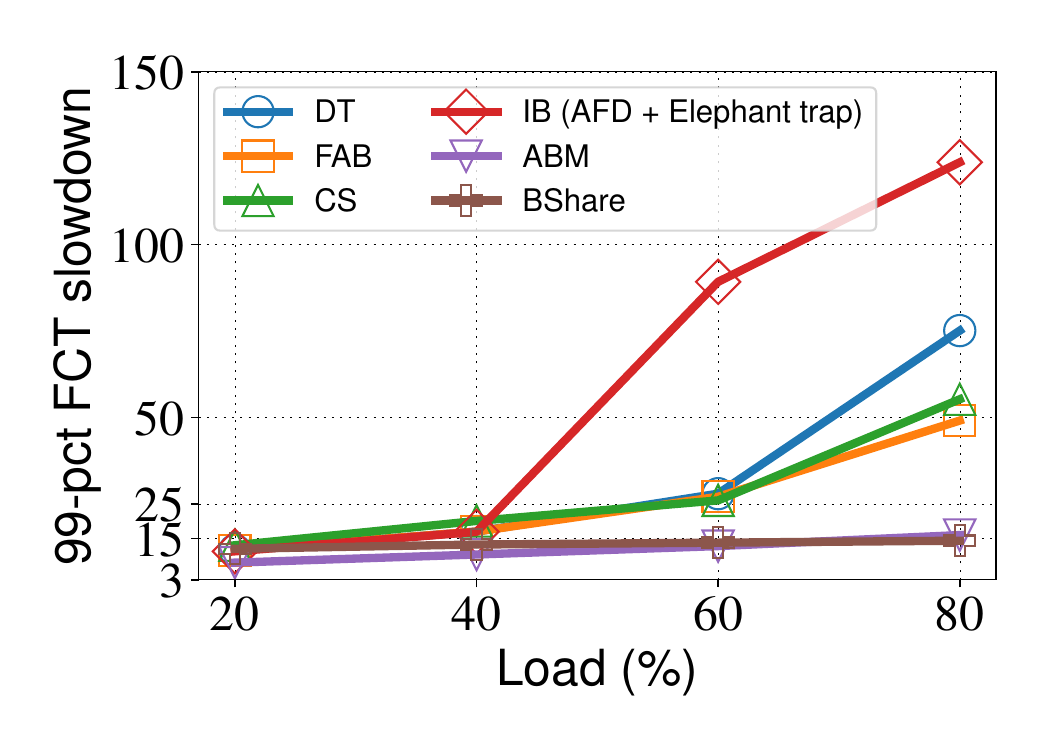}
        \caption{\texttt{Short Flows}}
        \label{fig:load:Short}
    \end{subfigure}    
    \begin{subfigure}[b]{.24\linewidth}
        \centering
        \includegraphics[width=\linewidth]{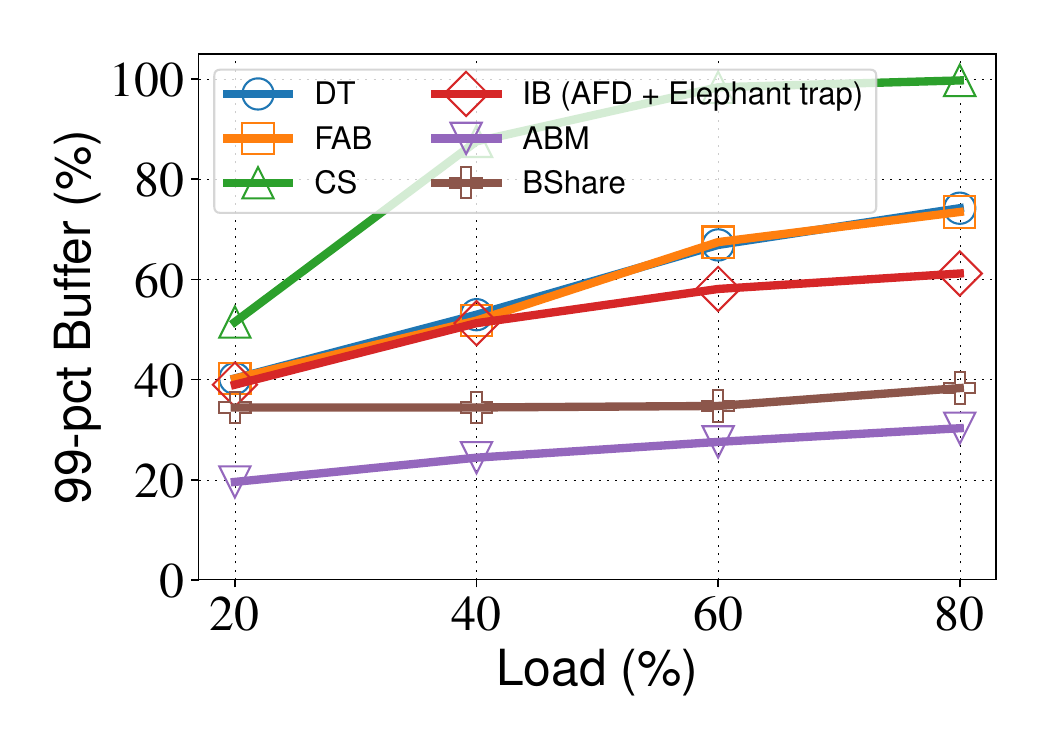}
        \caption{\texttt{Buffer Occupancy}}
        \label{fig:load:Buffer}
    \end{subfigure}
    \begin{subfigure}[b]{.24\linewidth}
        \centering
        \includegraphics[width=\linewidth]{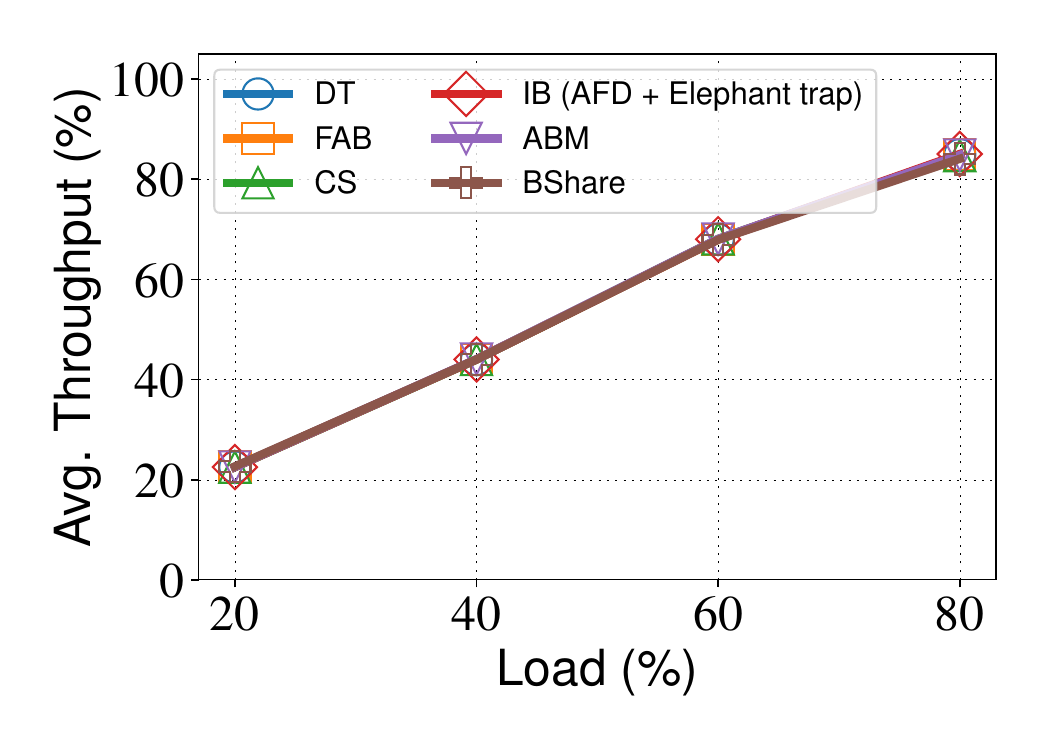}
        \caption{\texttt{Throughput}}
        \label{fig:load:Throughput}
    \end{subfigure}    
    \vspace{-5pt}
    \caption{Comparison of buffer management (BM) schemes under varying network loads. \systemname achieves performance comparable to ABM and outperforms other BM schemes in key scenarios:
    (a) incast flows that generate bursty traffic,
    (b) short flows representative of web search workloads.
    In addition, \systemname
    achieves lower buffer occupancy (c), and
     sustains high throughput without degradation (d).}
    \label{fig:loadVsFCT}
\end{figure*}

We simulate \systemname through packet-level simulations in NS-3 and compare it against state-of-the-art buffer management schemes tailored for datacenter networks. 
Our simulations use a leaf-spine datacenter topology comprising eight spine switches and 256 servers connected via eight leaf switches. All links operate at 10Gbps and incur a 10us propagation delay. The network follows a 4:1 oversubscription ratio, mirroring the configuration used in prior work~\cite{abm-sigcomm22}. Switches in both the leaf and spine layers are provided with 9.6KB of buffer per port per Gbps, consistent with the characteristics of the Broadcom Trident II ASIC~\cite{ShallowBuffer-ToN21,trident2}.

\medskip
We evaluate \systemname under two representative datacenter traffic patterns. First, we use the web search workload from~\cite{dctcp-sigcomm10}, which captures realistic flow size distributions observed in production datacenters. This workload is used to evaluate performance across a range of network loads, varying from 20\% to 80\%.

Second, we generate incast traffic patterns, following the methodology in~\cite{abm-sigcomm22}, to emulate bursty query-response behavior typical of distributed storage systems. In each incast scenario, a single query triggers simultaneous responses from multiple servers. We configure each server to issue two requests per second, and we vary the total burst size from 10\% to 100\% of the switch’s buffer capacity to stress the system under varying degrees of congestion.

To evaluate performance under different congestion control strategies, we run experiments using two widely studied transport protocols: DCTCP~\cite{dctcp-sigcomm10} and PowerTCP~\cite{powertcp-nsdi22}. These protocols play a crucial role in how data is transmitted and managed across the network, allowing us to evaluate their performance under different traffic conditions.

\medskip
\noindent\textbf{Benchmarks.} 
We compare \systemname with five representative BM schemes drawn from both academic literature and industry practice: Active Buffer Management (ABM)~\cite{abm-sigcomm22}, Dynamic Thresholds (DT)~\cite{DT-ToN98}, Flow-Aware Buffering (FAB)\cite{Flowaware-buffer19}, Complete Sharing (CS), and Cisco Intelligent Buffering (IB)~\cite{CIB}.
Across the board, ABM allocates buffer space based on the normalized drain rate and the proportion of remaining buffer space. DT uses a simpler approach, allocating buffer strictly according to remaining capacity. FAB extends DT by incorporating flow-awareness, giving preferential treatment to short flows. CS allows any queue to grow as long as buffer space is available, emulating an aggressive complete sharing policy. IB combines DT with approximate fair dropping through a hierarchical scheme, reflecting buffer management heuristics deployed in commercial Cisco switches.

To understand how \systemname interacts with different congestion control strategies, we evaluate it alongside several transport-layer protocols: Cubic~\cite{Cubic} (loss-based), DCTCP~\cite{dctcp-sigcomm10} (ECN-based), TIMELY~\cite{Timely-sigcomm15} (RTT-gradient-based), PowerTCP, and $\theta$-PowerTCP~\cite{powertcp-nsdi22} (power-aware). Due to space constraints, we present representative results that capture the key trends across these protocols.

We report on three primary performance metrics: total switch buffer occupancy, throughput, and FCT slowdown. We calculate FCT slowdown as the ratio between the actual FCT and the ideal FCT measured in the absence of competing traffic, providing a normalized view of flow latency under load.

\begin{figure*}[t]
    \centering    
    \begin{subfigure}[b]{.24\linewidth}
        \centering
        \includegraphics[width=\linewidth]{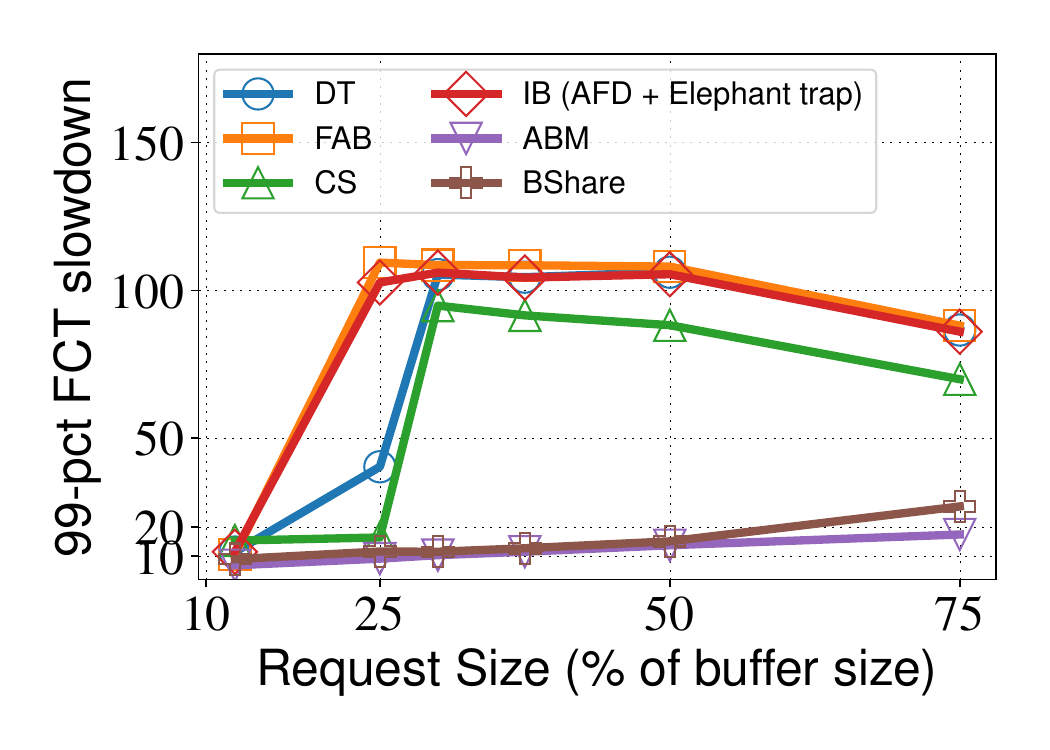}
        \caption{\texttt{Incast Flows}}
        \label{fig:burst:Incast}
    \end{subfigure}    
    \hfill
    \begin{subfigure}[b]{.24\linewidth}
        \centering
        \includegraphics[width=\linewidth]{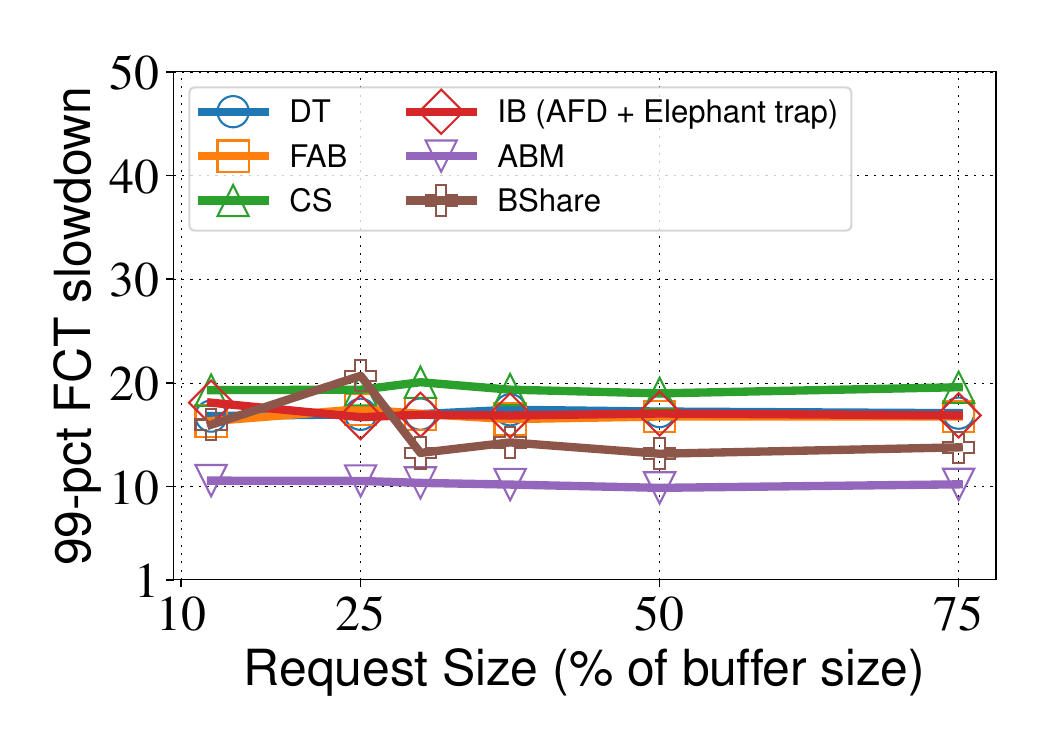}
        \caption{\texttt{Short Flows}}
        \label{fig:burst:Short}
    \end{subfigure}    
    \begin{subfigure}[b]{.24\linewidth}
        \centering
        \includegraphics[width=\linewidth]{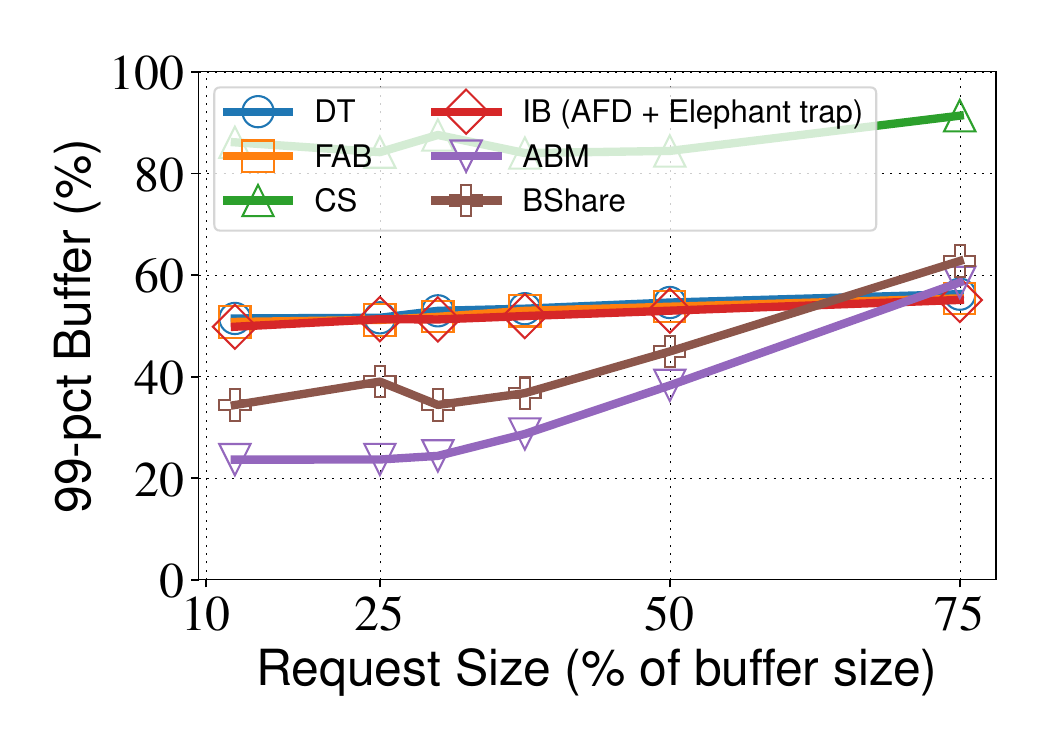}
        \caption{\texttt{Buffer Occupancy}}
        \label{fig:burst:Buffer}
    \end{subfigure}
    \begin{subfigure}[b]{.24\linewidth}
        \centering
        \includegraphics[width=\linewidth]{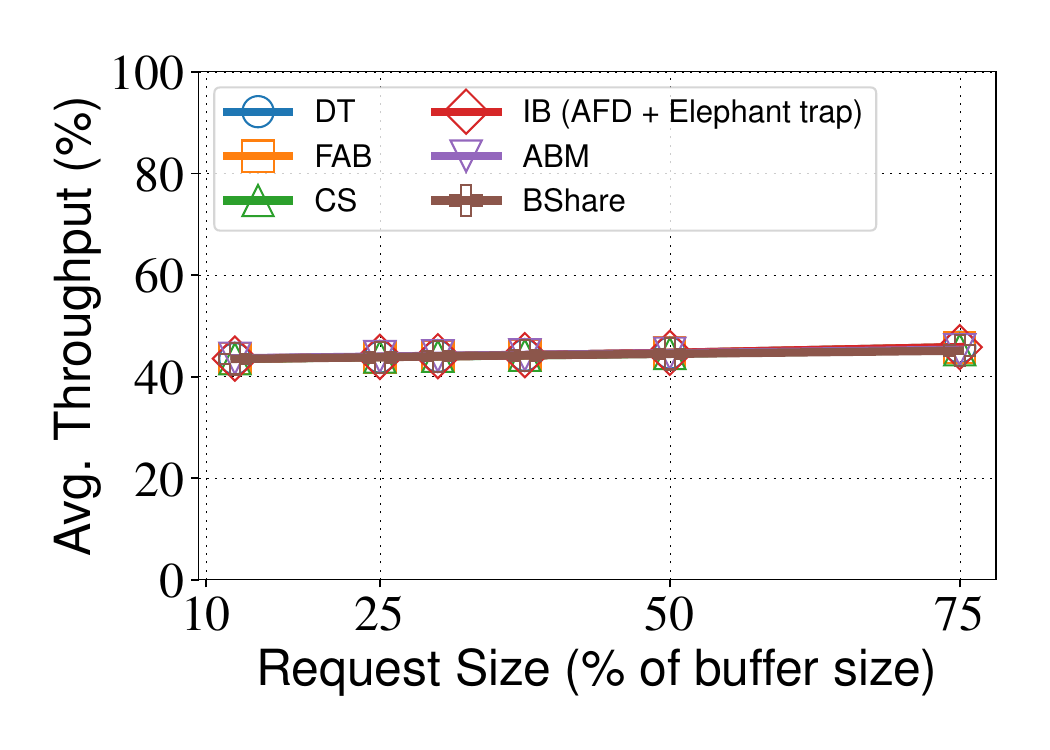}
        \caption{\texttt{Throughput}}
        \label{fig:burst:Throughput}
    \end{subfigure}    
    \vspace{-5pt}
    \caption{Comparison of BM schemes under various bursts size. \systemname demonstrates comparable performance to ABM and superior performance compared to other BM schemes:  (a) for flows contributing to bursts (incast traffic), and (b) for short flows (websearch), across different bursts. Additionally, \systemname  utilizes less buffer (c), and  maintains throughput  (d).}
    \label{fig:burstVsFCT}
\end{figure*}

\noindent\textbf{Parameters configuration.} 
We set $\alpha$ to 0.5 for \systemname, ABM, DT, and FAB. For DCTCP, PowerTCP, and $\theta$-PowerTCP parameters were set as per~\cite{dctcp-sigcomm10,powertcp-nsdi22}. We update the values of $c_{p}$ and the ABM drain rate once per RTT.

\subsubsection{Incast Traffic FCT's}
\systemname significantly reduces the 99th percentile FCT slowdown in incast scenarios compared to traditional buffer management schemes. Figure~\ref{fig:load:Incast} presents results under Cubic transport with a fixed request size equal to 30\% of the switch buffer, across varying network loads. At low load (20\%), \systemname performs similarly to existing schemes, including DT, FAB, CS, IB, and ABM. However, as load increases, \systemname consistently outperforms DT, FAB, CS, and IB, and closely tracks ABM's performance.
At 40\% load, \systemname reduces FCT slowdown by an average of 88.86\% compared to DT, FAB, CS, and IB. At 80\% load, this gap further widens, with a 93.7\% reduction in FCT slowdown, while still maintaining performance comparable to ABM. These results highlight \systemname's robustness under high-incast pressure and demonstrate its effectiveness in preserving latency-sensitive performance, especially under aggressive buffer contention.

Figure~\ref{fig:burst:Incast} shows the 99th percentile FCT slowdown for incast flows under a fixed 40\% web search workload, as the incast request size varies. Even with small bursts—e.g., 12.5\% of the buffer size—\systemname delivers a meaningful reduction in FCT slowdown. On average, it improves FCT slowdown by 25.85\% compared to DT, FAB, CS, and IB, and performs comparably to ABM.
As the size of the request increases, the benefits of \systemname become more apparent. For example, at a request size of 50\% of the buffer size, \systemname shows a remarkable reduction in the FCT slowdown for the incast workload, averaging at least \text{85.21\%} compared to DT, FAB, CS, and IB, and again close to ABM.

The results reveal that as network load increases, \systemname consistently outperforms traditional buffer management schemes, including DT, FAB, CS, and IB, and closely approaches the performance of ABM in all evaluated scenarios.

\begin{figure*}[tp]
    \centering    
    \begin{subfigure}[b]{.24\linewidth}
        \centering
        \includegraphics[width=\linewidth]{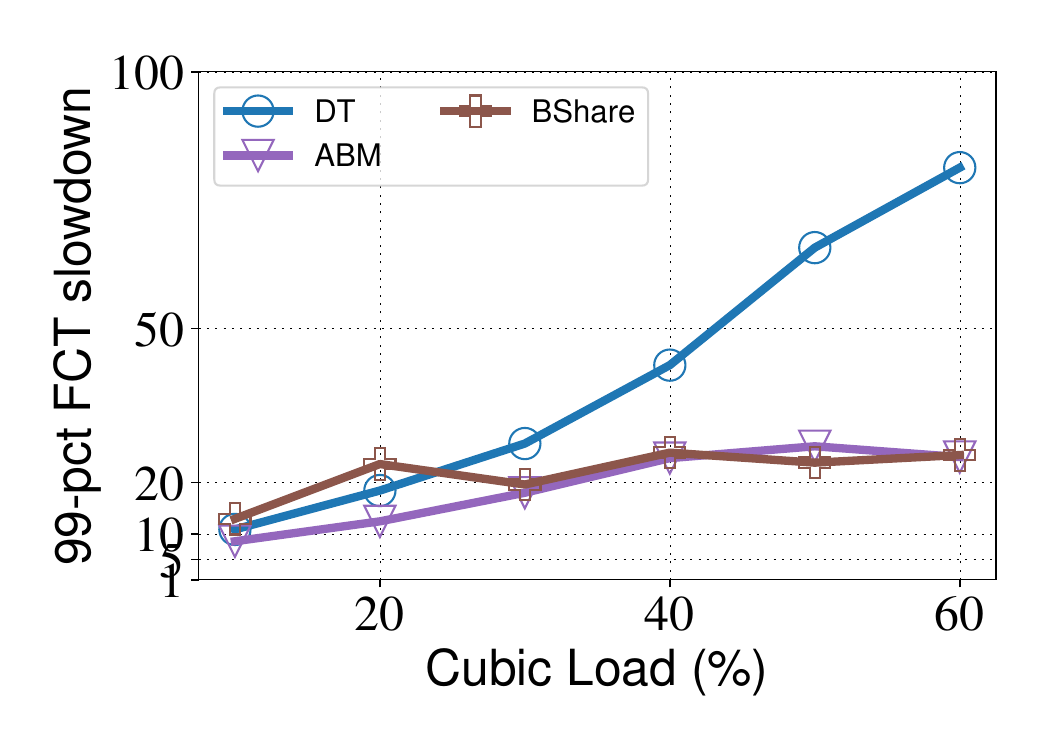}
        \caption{\texttt{Cubic Flows}}
        \label{fig:multi:Cubic}
    \end{subfigure}    
    \hfill
    \begin{subfigure}[b]{.24\linewidth}
        \centering
        \includegraphics[width=\linewidth]{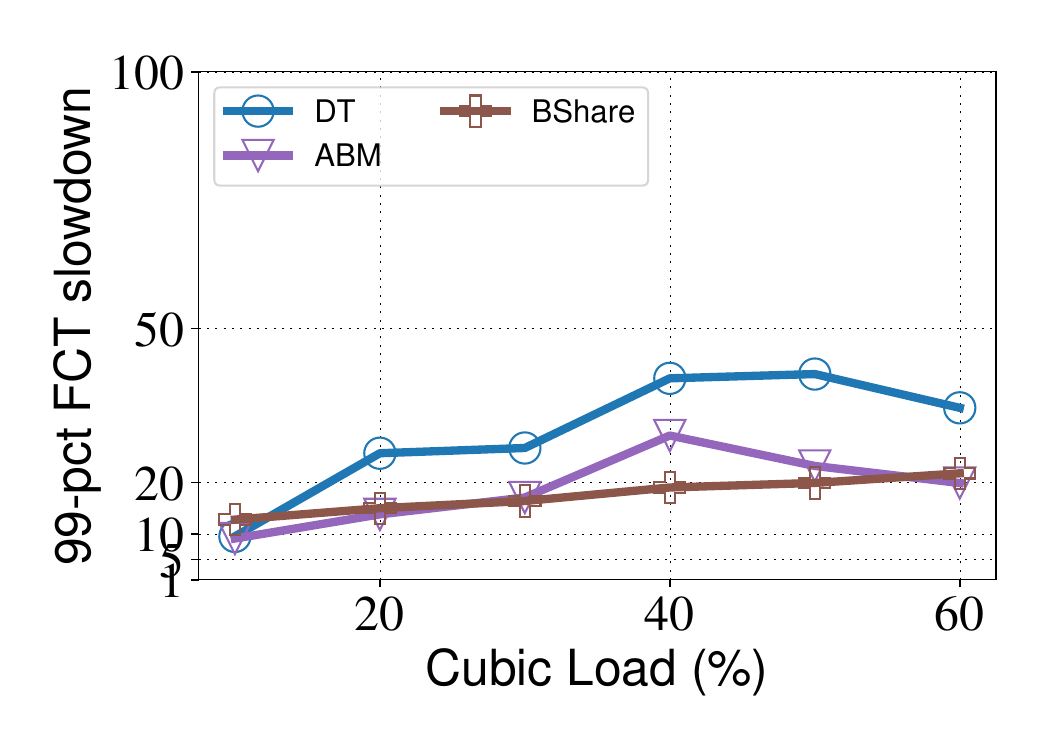}
        \caption{\texttt{DCTCP Flows}}
        \label{fig:multi:DCTCP}
    \end{subfigure}    
    \begin{subfigure}[b]{.24\linewidth}
        \centering
        \includegraphics[width=\linewidth]{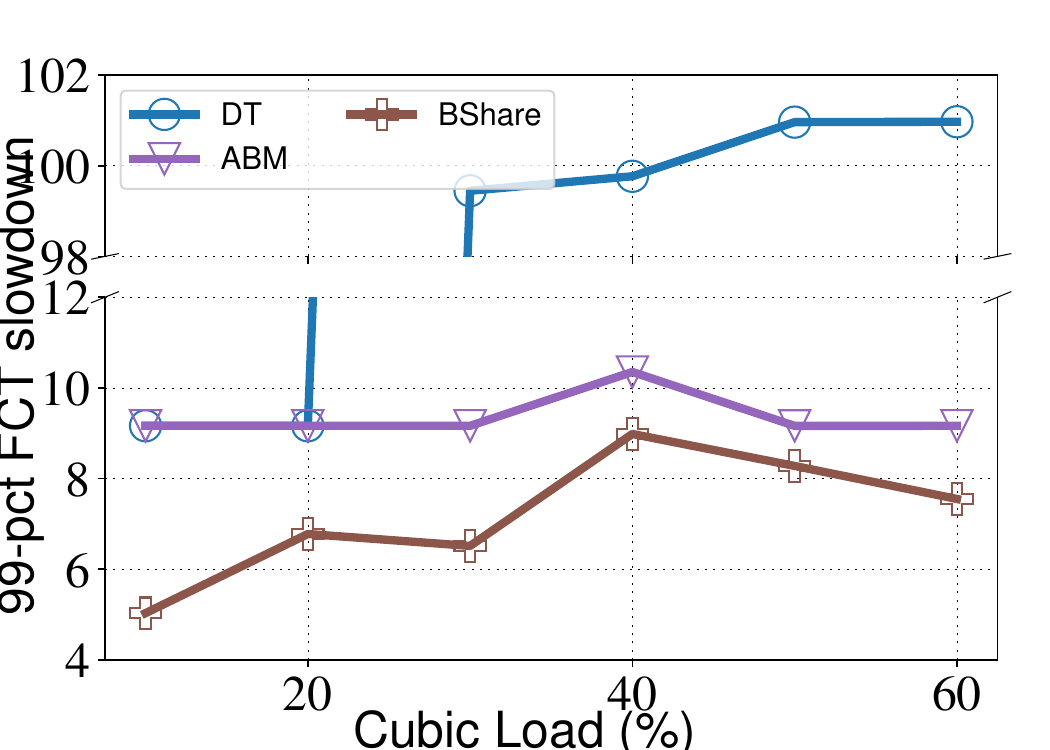}
        \caption{\texttt{$\theta$-PowerTCP}}
        \label{fig:multi:thetapower}
    \end{subfigure}
    \begin{subfigure}[b]{.24\linewidth}
        \centering
        \includegraphics[width=\linewidth]{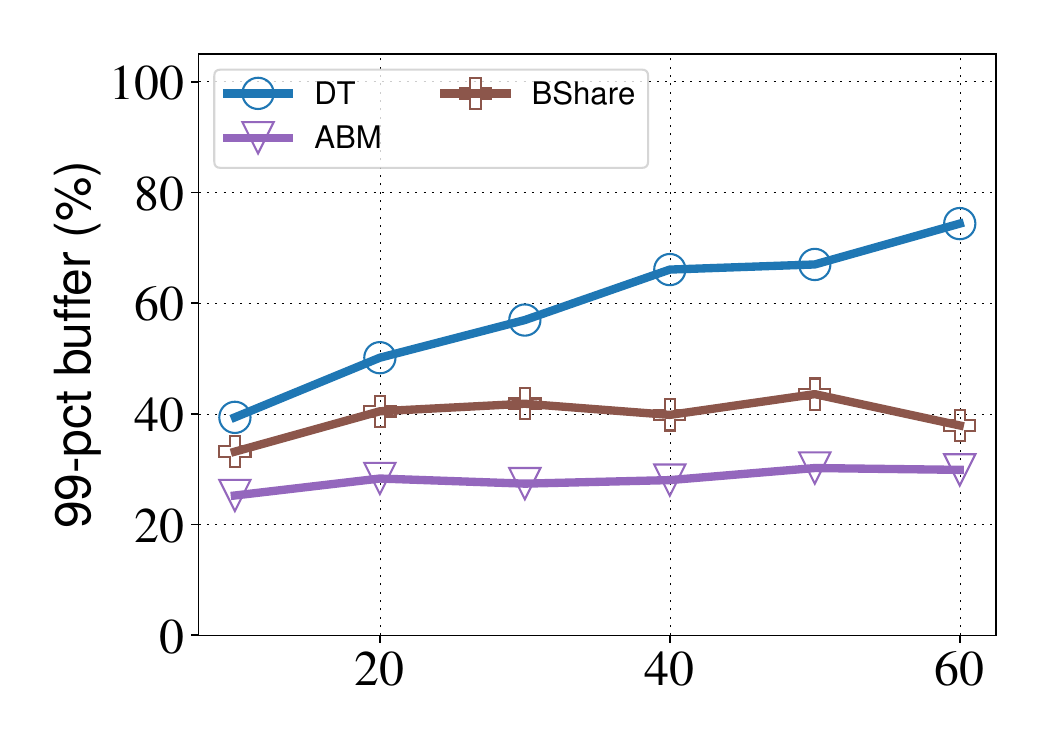}
        \caption{\texttt{Buffer occupancy}}
        \label{fig:multi:buffer}
    \end{subfigure}    
    \vspace{-5pt}
    \caption{The behaviour of Cubic, DCTCP, and $\theta$-PowerTCP when flows use separate queues, each managed by one of three BM algorithms: DT, ABM, and \systemname. \systemname and ABM achieve a similar FCT slowdown while \systemname uses slightly more buffer space.}
    \label{fig:multiVsFCT}
\end{figure*}
\subsubsection{Buffer Utilization}

\systemname efficiently allocates the available buffer space of the switch for medium and long flows, even under buffer-intensive transport protocols such as Cubic. As shown in Figure~\ref{fig:load:Buffer}, it reduces average buffer occupancy by 41.5\% across varying loads compared to traditional BM schemes, while maintaining throughput comparable to ABM.

In addition to overall efficiency, \systemname adapts buffer usage to accommodate bursty traffic. Figure~\ref{fig:burst:Buffer} illustrates that as the incast request size increases, \systemname strategically utilizes more buffer to absorb bursts. For large request sizes, \systemname consumes 12.92\% more buffer than DT, FAB, and IB, and 7.18\% more than ABM, highlighting its ability to dynamically allocate resources when needed without compromising performance.

\systemname reduces average buffer utilization compared to traditional schemes while dynamically allocating additional buffer during burst events--demonstrating both efficiency and responsiveness. In high-burst scenarios, it surpasses ABM in buffer usage to better absorb transient congestion.

\subsubsection{Throughput}
We evaluate \systemname's throughput performance under both varying network loads and incast burst sizes. As shown in Figure~\ref{fig:load:Throughput}, \systemname sustains throughput comparable to existing buffer management schemes across all load levels. Even under high burst pressure—illustrated in Figure~\ref{fig:burst:Throughput} with large incast request sizes—\systemname maintains stable throughput while continuing to deliver low FCTs for short flows.

\systemname preserves throughput on par with state-of-the-art schemes, while upholding delay-based thresholds and achieving low tail latency for short flows.

\subsubsection{Distinct Priorities}
We evaluate the performance of \systemname across distinct priority classes using a scenario in which Cubic, DCTCP, and $\theta$-PowerTCP flows are each assigned separate queues for both web search and incast workloads. Figure~\ref{fig:multiVsFCT} shows that \systemname significantly improves FCT slowdown compared to DT and ABM under varying load conditions.

For Cubic flows, as load increases, \systemname maintains competitive performance and begins to outperform ABM. At 50\% load, it reduces FCT slowdown by 11.80\% relative to ABM and by 63.72\% relative to DT. In the case of $\theta$-PowerTCP, \systemname consistently outperforms both baselines--achieving a 45.07\% reduction in FCT slowdown at 10\% load and 28.89\% at 30\% load compared to ABM, and over 93.44\% compared to DT.

For DCTCP flows, \systemname continues to show strong performance, reducing FCT slowdown by up to 34.73\% over ABM and 52.79\% over DT. These improvements highlight \systemname's ability to prioritize flows effectively across multiple congestion control algorithms and traffic types.

In distinct priority scenarios, 
\systemname consistently outperforms DT and often surpasses ABM, with particularly strong gains under $\theta$-PowerTCP across all load levels.

\subsubsection{Realistic Buffer}

We evaluate \systemname under realistic buffer constraints, focusing on a baseline of 9.6\textit{KB} of buffer per port per Gbps, representative of Broadcom Trident II switches~\cite{trident2}. To assess generality and hardware applicability, we also simulate smaller, implementable buffer sizes as seen in Tomahawk and Tofino switches~\cite{ShallowBuffer-ToN21}.

We consider both web search and incast workloads at 40\% network load, fixing the incast request size to 25\% of the buffer size defined for Trident II. Evaluations are conducted using DCTCP and PowerTCP transport protocols.

\begin{figure}[tp]
    \centering    
    \begin{subfigure}[b]{.48\linewidth}
        \centering
        \includegraphics[width=\linewidth]{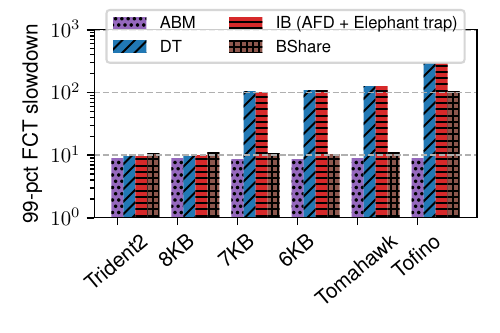}
        \caption{\texttt{DCTCP}}
        \label{fig:buf:DCTCP}
    \end{subfigure}    
    \hfill
    \begin{subfigure}[b]{.48\linewidth}
        \centering
        \includegraphics[width=\linewidth]{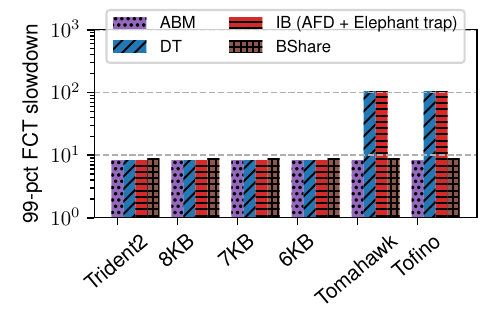}
        \caption{\texttt{PowerTCP}}
        \label{fig:buf:PowerTCP}
    \end{subfigure}      
    \vspace{-5pt}
    \caption{Comparison of buffer management (BM) schemes under realistic buffer sizes across Trident II, Tomahawk, and Tofino switch configurations. \systemname consistently maintains low 99th percentile FCT across buffer sizes, with the exception of a slight degradation on Tofino under DCTCP. Under PowerTCP, \systemname achieves lower tail latency than DT and IB, and performs on par with ABM.}
    \label{fig:buffer}
\end{figure}

Figure~\ref{fig:buffer} presents the 99th percentile FCT slowdown for incast flows under different buffer sizes. Under DCTCP (Figure~\ref{fig:buf:DCTCP}), \systemname maintains stable performance across all configurations, with only a minor increase in FCT slowdown observed on Tofino. 
In contrast, both DT and IB degrade sharply when buffer sizes drop below 7KB per port per Gbps--showing up to a 10$\times$ increase in FCT slowdown relative to \systemname, particularly with Tomahawk and DCTCP.

Under PowerTCP (Figure~\ref{fig:buf:PowerTCP}), DT and IB appear more resilient until buffer sizes fall to 6KB per port per Gbps. However, their limitations become evident at 5.12KB, where both schemes fail to maintain low tail latencies. \systemname, by contrast, continues to deliver near-ABM performance across all buffer sizes tested.

Across a range of realistic hardware buffer sizes, \systemname closely tracks ABM's performance under both DCTCP and PowerTCP. It remains robust even when buffer capacity is severely constrained, where other schemes like DT and IB suffer substantial tail latency degradation.

\section{Related Work}
Switch buffer management has been widely studied in the literature via different approaches. We categorized them into hardware-based, BM, AQM, and hybrid techniques.

\noindent\textbf{Hardware-based.}
Hardware-based solutions aim to manage the buffer directly in the switch. However, they often require additional hardware support and may have limitations in handling asymmetric traffic. For instance, R-ACK~\cite{RACK-ToN22} and ABQ~\cite{ABQ} offer innovative approaches to control transmission rates and manage ACK packets, but their implementation complexity and dependency on hardware support pose challenges.

\noindent\textbf{BM-based.}
BM algorithms dynamically allocate buffer space among different queues or flows within network devices to prevent congestion and ensure fair distribution. Example algorithms are PO~\cite{OP,OPBM}, DT~\cite{DT-ToN98}, and EDT~\cite{EDT-TPDS17} that aim to excel in buffer space efficiency and burst absorption. However, implementing some BM algorithms, like PO, can be complex and may require additional hardware support.

\noindent\textbf{AQM-based.}
AQM algorithms focus on regulating queue lengths by selectively admitting or dropping incoming packets based on predefined thresholds. The AQM schemes like PIE~\cite{PIE-hpsr13}, CoDel~\cite{CoDel}, and RED~\cite{RED-ToN93} effectively handle bursty traffic and aim to maintain low queue lengths to prevent congestion. However, they may face challenges in switch fabric implementation due to their reliance on calculating average queue length and drain rate.

\noindent\textbf{Hybrid-based.}
ABM~\cite{abm-sigcomm22} combines the strengths of BM and AQM algorithms to achieve robust isolation properties and stable buffer drain times. ABM offers efficient burst absorption and high predictability while maintaining high throughput. Although ABM exhibits promising performance improvements over existing techniques, calculating the drain rate for each queue remains challenging, particularly in fluctuating link capacities. Unlike ABM, \systemname does not need the drain rate calculation. L2BM~\cite{l2bm-icdcs23} has a similar logic to ABM. Credence~\cite{credence-nsdi24} proposes a drop-tail buffer sharing algorithm with machine learning-based prediction to improve performance.

\section{Conclusion}
In this paper, we presented \systemname, a practical and delay-aware buffer sharing mechanism for datacenter switches, inspired by CoDel's queuing delay signals. \systemname is designed to address the challenges of dynamic buffer allocation in shared-memory switches using a lightweight, single-parameter control policy. Unlike prior approaches that rely on complex rate estimation or multi-stage coordination, \systemname requires no predictive modeling.
Through extensive simulation, we show that \systemname consistently reduces flow completion times across a range of traffic loads, buffer sizes, and transport protocols—including under burst-heavy workloads and with advanced congestion control like PowerTCP.

\label{LastPage}

\balance
\bibliographystyle{plain}

\bibliography{reference}


\end{document}